\documentclass[prA,twocolumn,floatfix,amsmath]{revtex4}

\usepackage{amsthm}
\usepackage{bbm}
\usepackage{amssymb}
\usepackage{amsmath}
\usepackage{graphicx}
\usepackage{epsfig}

\newtheorem*{thm*}{Theorem}

\theoremstyle{definition}

\theoremstyle{remark}

\newcommand{\sigm}{\tau}
\newcommand{\1}{\mathbbm{1}}

\DeclareMathOperator{\id}{id} \DeclareMathOperator{\trace}{tr}

\begin{document}

\title{Quantum Capacities of Channels with small Environment}

\author{Michael M. Wolf$^{1}$ and David P\'erez-Garc\'ia$^{2}$} \affiliation{$^{1}$ Max-Planck-Institute for Quantum
Optics, Hans-Kopfermann-Str.\ 1, D-85748 Garching,
Germany.\\$^{2}$ \'Area de Matem\'atica Aplicada, Universidad Rey
Juan Carlos, C/ Tulipan s/n, 28933 M\'ostoles (Madrid), Spain}

\begin{abstract}
We investigate the quantum capacity of noisy quantum channels
which can be represented by coupling a system to an effectively
small environment. A capacity formula is derived for all cases
where both system and environment are two-dimensional---including
all extremal qubit channels. Similarly, for channels acting on
higher dimensional systems we show that the capacity can be
determined if the channel arises from a sufficiently small
coupling to a qubit environment. Extensions to instances of
channels with larger environment are provided and it is shown that
bounds on the capacity with unconstrained environment can be
obtained from decompositions into channels with small environment.
\end{abstract}

\date{July 8, 2006}
\pacs{}

\maketitle

\section{Introduction}
One of the key concepts in both classical and quantum information
theory is the {\it capacity of a channel}. That is,
 the maximal number of bits---or qubits---that can be
transmitted reliably per use of the channel. In the classical
world Shannon's seminal coding theorem enables us to determine the
capacity of every classical channel. However, for the coherent
transmission of quantum information through a quantum channel no
comparable coding theorem is known.

In this work we determine the \emph{quantum capacity} for channels
which arise from interactions between a system and an effectively
small environment. Physically, these channels could correspond to
the dephasing of an electron spin in a quantum dot, the
spontaneous emission in a two-level atom or the loss of a photon
in an optical fiber. We will provide a capacity formula based on
the \emph{coherent information} for all cases where system and
environment are qubits. This includes in particular all extremal
qubit channels, i.e., those into which any qubit channel can be
decomposed. For larger systems we find a similar result if only
the coupling to the environment is sufficiently small. Note that
the size of the environment which matters in this context is not
the physical one, but the effective size given by a minimal
representation of the channel. Along the way we find instances of
channels with larger environment to which the results can be
extended. Moreover, we show that additivity of the quantum
capacity implies its convexity so that upper bounds on the quantum
capacity of channels with unconstrained environment can be
obtained from mixing channels with small environment.

We will start by introducing the basic tools and notions, then
derive the capacity formula in the qubit case and finally treat
systems of higher dimension.

\section{Preliminaries}

Consider a quantum system characterized by a density operator
$\rho$ of dimension $d$. Every quantum channel $T$ can be
represented by a unitary coupling to an environment which is
initially in a pure state $\varphi_E$ of dimension $d_E\leq d^2$.
That is, $\rho\mapsto
T(\rho)=\mbox{tr}_E\big[U(\rho\otimes\varphi_E)U^\dagger\big]$.
Alternatively we can write any channel as
$T(\rho)=\sum_{i=1}^{d_E} A_i\rho A_i^\dagger$ where the
\emph{Kraus operators} $A_i$ fulfill  $\sum_iA_i^\dagger A_i=\1$.
The \emph{conjugate channel}
$\tilde{T}(\rho)=\mbox{tr}_S\big[U(\rho\otimes\varphi_E)U^\dagger\big]$
is defined as a mapping from the system (which is traced out in
the end) to the environment. Its Kraus operators are given by
$(\tilde{A}_i)_{kl}=(A_k)_{il}$ \cite{Beth2}.

The \emph{quantum capacity} $Q(T)$ is the maximal asymptotically
achievable rate at which we can reliably transmit quantum
information---measured in number of qubits---through a channel
(cf.\cite{KW}). A major theoretical step was the proof of the
\emph{capacity theorem} \cite{Qcap} stating that
\begin{eqnarray}
Q(T) &=& \lim_{n\rightarrow\infty}\frac1n\; \sup_\rho
J\big(T^{\otimes n},\rho\big)\;,\label{Qreg}\\
J\big(T,\rho\big) &=&
S\big(T(\rho)\big)-S\big(\tilde{T}(\rho)\big)\;,\label{J}
\end{eqnarray} where $J$ is called \emph{coherent information} and
$S(\rho)=-\trace[{\rho\log_2\rho}]$ is the von Neumann entropy.
The evaluation of the  expression in Eq.(\ref{Qreg}) is a daunting
task. First, the regularization $n\rightarrow\infty$ appears to be
generally necessary since there are channels for which the
maximized coherent information is non-additive \cite{nonadditive}.
Second, $J(T,\rho)$ is in general not concave  in $\rho$ allowing
for a complex landscape of local maxima which are not global ones.
We will show in the following that for the considered channels
with small environment these two obstacles can, however, be
avoided so that the calculation of $Q(T)$ becomes a feasible task.
The main tool behind is the concept of \emph{degradability} of a
channel introduced by Devetak and Shor \cite{DS}.

A channel is called \emph{degradable} if it can simulate its
conjugate in the sense that there is another channel $\Phi$ which
composed with $T$ yields $\tilde{T}=\Phi\circ T $. Similarly, we
call it \emph{anti-degradable} if the conjugate $\tilde{T}$ is
degradable. The importance of this property stems from the fact
that the coherent information then becomes a conditional entropy
\cite{DS} which is in turn sub-additive and concave. In other
words, for a set of degradable channels $T_i$ one has
\begin{equation}
Q\Big(\bigotimes_i T_i\Big)=\sum_i\sup_\rho J(T_i,\rho)\;,
\end{equation}
for which  local maxima are already global ones. On the other
hand, if $T$ is anti-degradable then the no-cloning theorem
implies $Q(T)=0$. Channels which are known to be (anti-)degradable
are lossy bosonic channels \cite{CG,WGPG}, channels with diagonal
Kraus operators \cite{DS} and qubit amplitude damping channels
\cite{GF}.

Let us now discuss how to check whether a channel is
(anti-)degradable. To this end consider the generic case where $T$
is invertible. Then degradability is equivalent to complete
positivity of $\Phi=\tilde{T}\circ T^{-1}$. Similarly,
anti-degradability is related to complete positivity of
$\Phi^{-1}$. To express this in a more convenient form we exploit
Jamiolkowski's operator-map duality \cite{Jami} which assigns a
bipartite operator $\sigm=\big(T\otimes\id\big)(\omega)$ to each
channel $T$ by sending half of a (unnormalized) maximally
entangled state $\omega=\sum_{i,j=1}^d|ii\rangle\langle jj|$
through $T$. Similarly, we can assign to each channel a
\emph{transfer matrix} $\sigm^\Gamma$
 which is obtained via the involution $\langle
ij|\sigm^\Gamma|kl\rangle=\langle ik|\sigm|jl\rangle$. The
advantage of these two representations is that complete positivity
of $T$ reduces to $\sigm\geq 0$ and concatenating and inverting
channels boils down to matrix multiplication and matrix inversion
on the level of $\sigm^\Gamma$. Hence, checking degradability
becomes equivalent to verifying positivity of the eigenvalues of
\begin{equation}\label{DegCond}
\sigm_{\Phi}=\Big[{\tilde{\sigm}}^\Gamma
\big(\sigm^\Gamma\big)^{-1}\Big]^\Gamma\geq 0\;.
\end{equation}
Likewise, anti-degradability amounts to positivity of
$\sigm_{\Phi^{-1}}$ and if both
$\sigm_{\Phi},\sigm_{\Phi^{-1}}\geq 0$ then $T$ and $\tilde T$ are
unitarily equivalent. We will now apply these tools to channels
with small environment.

\section{Qubit channels}

Consider qubit channels with a qubit environment, i.e., $d=d_E=2$.
We will first show that every such channel is either degradable or
anti-degradable, then derive a capacity formula and finally sketch
the application of the result to arbitrary qubit channels. The
latter will be based on a general convexity property for the
quantum capacity.

\begin{figure}[t]
\begin{center}
\epsfig{file=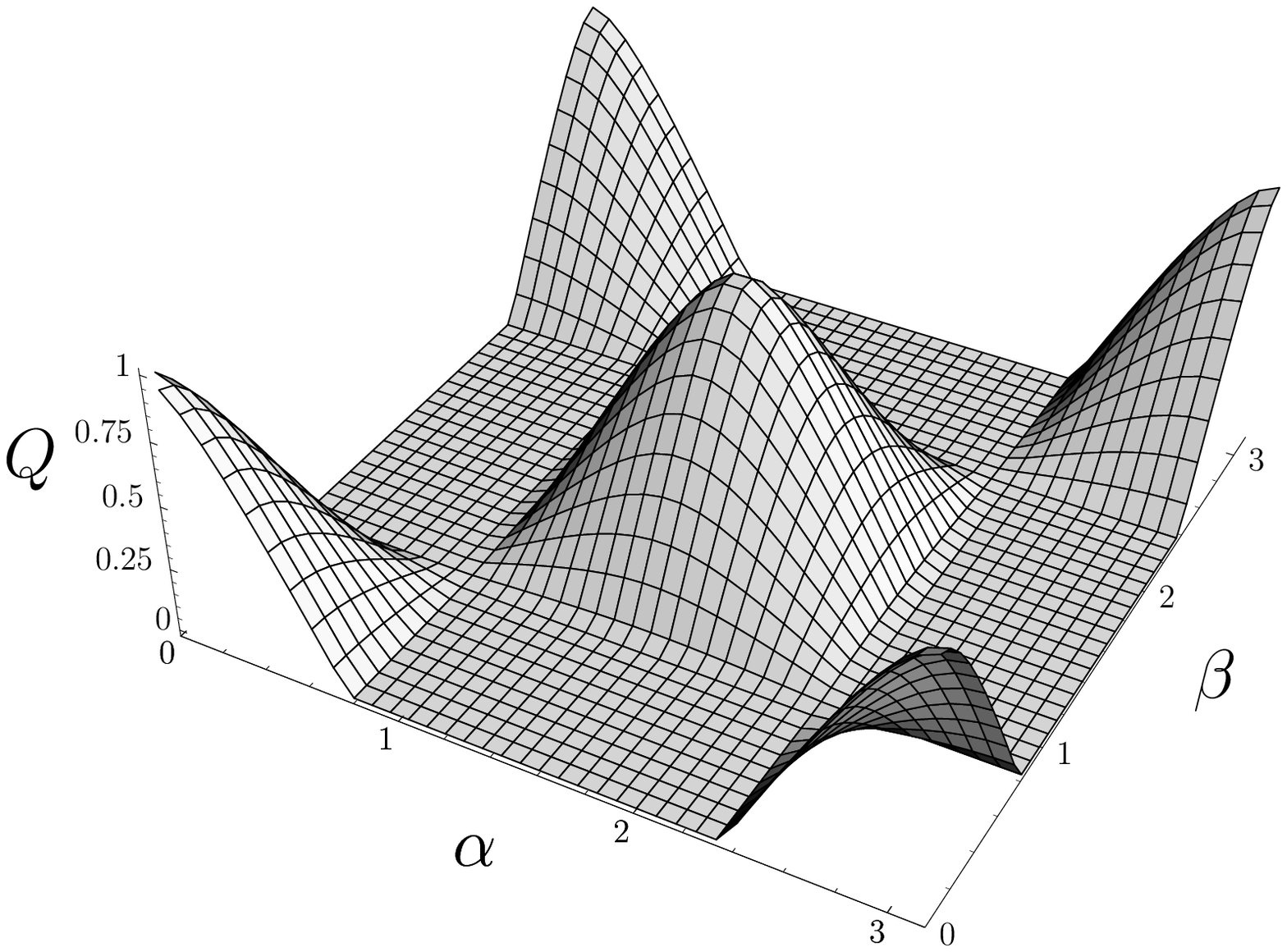,angle=0,width=0.84\linewidth}
\end{center}
\caption {Quantum capacity $Q$ of extremal qubit channels
parameterized by the normal form in Eq.(\ref{A2}). For
$\alpha=\beta$ this represents a \emph{dephasing channel}
 and for $\beta=0$ an \emph{amplitude damping channel}. Whereas the peaks $Q=1$ reflect  unitary evolutions, the regions with $Q=0$
 correspond to anti-degradable channels for which most of the information is lost to the environment.} \label{figarray}
\end{figure}

To start with we utilize that two channels have the same capacity
if they differ merely by unitaries acting on input and output.
Building  equivalence classes in this way reduces the number of
parameters to two ($\alpha,\beta\in\mathbb{R}$) and allows us
(following \cite{Ruskai}) to represent every such channel by Kraus
operators in the normal form
\begin{equation}\label{A2}
A_1=\left(%
\begin{array}{cc}
  \cos\alpha & 0 \\
  0 & \cos\beta \\
\end{array}%
\right),\quad A_2=\left(%
\begin{array}{cc}
  0 & \sin\beta \\
  \sin\alpha & 0 \\
\end{array}%
\right)\;.
\end{equation}
For $\alpha=\beta$ this represents a \emph{dephasing channel}
\cite{DS} and for $\beta=0$ an \emph{amplitude damping channel}
\cite{GF}. As the set of invertible channels $T$ (and conjugates
$\tilde{T}$) is dense, it is sufficient to consider the invertible
case---the general statement will then follow from continuity
\cite{cont}. We denote the spectra
 which we have to check according to Eq.(\ref{DegCond}) by
$\mbox{spec}(\sigm_\Phi)=\{0,0,\lambda_1,\lambda_2\}$ and
$\mbox{spec}(\sigm_{\Phi^{-1}})=\{0,0,\tilde\lambda_1,\tilde\lambda_2\}$.
The reason for the two-dimensional null space will become clear
below Eq.(\ref{H}). Straight forward algebra shows that
\begin{equation}\label{lambdas}
\frac{\lambda_1}{\lambda_2}=-\frac{\tilde\lambda_1}{\tilde\lambda_2}=\frac{\cos
2\alpha}{\cos 2\beta}\;.
\end{equation}
As $\Phi$ and $\Phi^{-1}$ are trace preserving we have that
$\trace\sigm_\Phi=\trace\sigm_{\Phi^{-1}}=d$ so that in both cases
at most one of the eigenvalues can be negative. Together with
Eq.(\ref{lambdas}) this implies that \emph{either} $\sigm_\Phi\geq
0$ \emph{or} $\sigm_{\Phi^{-1}}\geq 0$. Hence, the capacity $Q(T)$
is zero iff $\cos(2\alpha)/\cos(2\beta)\leq 0$ and given by the
supremum of the coherent information otherwise. In order to
evaluate the latter we note that the chosen representation in
Eq.(\ref{A2}) has the property of being covariant w.r.t. a
$\sigma_z$ Pauli rotation:
\begin{equation}
\sigma_z T\big( \rho\big)\sigma_z= T\big(\sigma_z
\rho\sigma_z\big),\quad \sigma_z \tilde{T}\big( \rho\big)\sigma_z=
\tilde{T}\big(\sigma_z \rho\sigma_z\big)\;.
\end{equation}
Together with Eq.(\ref{J}) and the concavity of the coherent
information for degradable channels this implies that
\begin{equation}
J\Big(T,(\rho+\sigma_z\rho\sigma_z)/2\Big)\geq
J\big(T,\rho\big)\;.
\end{equation}
Hence, diagonal input states maximize the coherent information and
joining pieces then yields the capacity formula (see Fig.1) in the
region of non-zero capacity ($\cos(2\alpha)/\cos(2\beta)>0$):
\begin{eqnarray}\label{Qformula}
Q(T)&=& \max_{p\in[0,1]}\;h\big(p \cos^2\alpha +(1-p)
\sin^2\beta\big)\\
&&\quad\quad\  - h\big(p \sin^2\alpha +(1-p)
\sin^2\beta\big)\;,\nonumber
\end{eqnarray}
where $h(x)=-x\log_2 x - (1-x)\log_2(1-x)$ is the binary entropy
function. This extends the findings of \cite{DS,GF} to arbitrary
qubit channels with two Kraus operators.

There are several ways of exploiting this result for qubit
channels with unconstrained environment. First, for any
composition $T=T_1\circ T_2$ of channels $T_i$ with known capacity
we can make use of the bottleneck inequality
$Q(T)\leq\min\{Q(T_1),Q(T_2)\}$. Second, we can consider convex
combinations $T=\sum_i p_i T_i$ with $p_i\geq 0$. In fact,
\emph{every} qubit channel can be convexly decomposed into
channels with $d_E=2$, as it was proven in \cite{Ruskai} that
every extremal qubit channel can be represented by Kraus operators
in the normal form of Eq.(\ref{A2}). In order to apply this we
need the following.

\section{Convexity of the Quantum Capacity}

The general behavior of the quantum capacity under mixing of
channels is not known---a problem reminiscent of a long-standing
question in the theory of entanglement distillation \cite{ED}.
However, one can easily show that additivity $Q\big(\bigotimes_i
T_i\big)=\sum_i Q(T_i)$ implies convexity $Q(\sum_i p_i T_i)\leq
\sum_i p_i Q(T_i)$. To see this we start from Eq.(\ref{Qreg})
applied to a convex combination $pT_1+(1-p)T_2$:
\begin{eqnarray}
&&\frac1n J\Big(\big(pT_1+(1-p)T_2\big)^{\otimes
n},\rho\Big)\nonumber\\ \nonumber
&\leq& \frac1n \sum_{m=0}^n\left(%
\begin{array}{c}
  n \\
  m \\
\end{array}%
\right) p^m (1-p)^{n-m} Q\big(T_1^{\otimes m}\otimes T_2^{\otimes
n-m}\big)\\
&=& p\; Q(T_1)+ (1-p)Q(T_2)\;,
\end{eqnarray}
where the inequality reflects convexity of the coherent
information w.r.t. the channel and the last step follows from the
additivity hypothesis together with the summation of the binomial
series. Note that the required additivity of $Q$ on tensor powers
is already implied by the simple additivity $Q(T_1\otimes
T_2)=Q(T_1)+Q(T_2)$ \cite{addQ}. Since the quantum capacity is
additive on degradable (resp. anti-degradable) channels, it is
also convex on these sets. That is, if we decompose a qubit
channel (now with unconstrained environment) into degradable
extremal channels, we get a simple upper bound on its capacity by
exploiting convexity together with Eq.(\ref{Qformula}). Moreover,
if a channel admits a convex decomposition into extremal channels
with zero capacity (i.e., anti-degradable ones), then it also has
vanishing capacity.

\section{Higher dimensions}

We will now investigate channels acting on higher dimensional
quantum systems. Our aim is to prove that a channel in any
dimension is degradable if it arises from a sufficiently small
coupling to a qubit environment. This will follow from a more
general result on channels which we call \emph{twisted diagonal}
and for which $d_E\leq d$.

Consider a completely positive map $T(\rho)=\sum_i A_i\rho A_i^{
\dagger}$ acting on a $d$ dimensional system with a $d_E$
dimensional environment. We will call $T$ \emph{twisted diagonal}
if there exist invertible matrices $X,Y$ such that $YA_i X$ is
diagonal with diagonal entries $a^{(i)}_l$, $l=1,\ldots,d$ and
associated normalized vectors $|\psi_l\rangle\propto
\sum_{i=1}^{d_E} a^{(i)}_l |i\rangle$. Related to these maps we
introduce a Hermitian $d\times d$ matrix $H$ with matrix elements
$H_{kl}=[(Y Y^\dagger)^{-1}]_{kl}/\langle\psi_k|\psi_l\rangle$ and
show that the positivity $H\geq 0 $ is equivalent to degradability
of $T$. To this end denote by $T_X, T_Y$ the completely positive
maps with one Kraus operator $X$ and $Y$ respectively. For
degradability we have to check complete positivity of $\tilde{T}
T^{-1}$. This is, however, equivalent to $\Psi=\tilde{T}T_X (T_Y T
T_X)^{-1}$ being completely positive, where the twisted diagonal
property now allows us to easily compute the inverse. The
Jamiolkowski operator $\sigm$ corresponding to $\Psi$ can then be
derived in a straight forward way and has the form
\begin{equation}\label{H} \sigm = \sum_{k,l=1}^d H_{kl}\;
|\psi_l\rangle\langle\psi_k|\otimes|l\rangle\langle k|.
\end{equation}
Since $\{|\psi_l\rangle\otimes|l\rangle\}$ is a set of orthonormal
vectors, $H$ and $\sigm$ have the same non-zero spectrum. Hence,
$H\geq 0$ iff $T$ is degradable. Moreover, in the special case of
diagonal channels for which $X=Y=\mathbbm{1}$ we get
$H=\mathbbm{1}$ recovering the result of \cite{DS} that all
diagonal channels are degradable.

The next step in our argumentation is to show that \emph{all}
channels with $d_E=2$ are twisted diagonal. More precisely, the
set of Kraus operators $\{A_1,A_2\}$ for which there exist
invertible matrices $X, Y$ such that $YA_iX$ is diagonal, is
dense. To see this consider the polar decomposition $A_i=U_iP_i$
and choose $Y=R P_1^{-1/2}U_1^\dagger$ and $X=P_1^{-1/2}R^{-1}$.
This maps $A_1\mapsto\mathbbm{1}$ and $A_2 \mapsto
R(P_1^{-1/2}U_1^\dagger U_2 P_2 P_1^{-1/2})R^{-1}$ so that we only
have to choose $R$ such that it diagonalizes the remaining part.

For channels arising from coupling a $d$-dimensional system to a
qubit environment we can thus resort to the degradability
criterion $H\geq 0$. In order to complete the argumentation we
note that $H=\mathbbm{1}$ for the ideal channel as well as for all
unitary evolutions of the system. As $H$ is continuous under
varying the channel, it will remain positive (and the channel thus
degradable) for a sufficiently small coupling to a qubit
environment. Hence, for all these channels one can efficiently
compute the quantum capacity.

 One might wonder how large this set of degradable channels is.
Clearly, for $d=2$ these are exactly half of the channels---the
other half is anti-degradable. Sampling channels according to the
Haar measure gives that 10\% (1\%) of the channels remain
degradable for $d=3\; (d=4)$.

\section{Conclusion}

In summary we showed that the quantum capacity can be calculated
for many channels arising from coupling a system to an effectively
small environment. This in particular completes the picture in the
case where both system and environment are qubits---analogous to
the Gaussian bosonic situation \cite{WGPG} where both are
characterized by a single mode. The results are based on the
\emph{degradability} of the considered channels. This property
fails to be true in general for larger environment---even in the
vicinity of the ideal channel \cite{SS}. However, one can use the
shown convexity property of the channel capacity (or that of
recently proposed assisted versions \cite{SSW}) in order to derive
upper bounds.

\section*{Acknowledgments}

We acknowledge valuable discussions with G. Giedke, K.G.H.
Vollbrecht and J.I. Cirac, and financial supported by
MTM2005-00082.


\begin{thebibliography}{99}
\bibitem{KW} D. Kretschmann, R.F. Werner, New J. Phys. {\bf 6}, 26 (2004).
\bibitem{Beth2} C. King, K. Matsumoto, M. Nathanson, M. B. Ruskai,
quant-ph/0509126.
\bibitem{Qcap} P.W. Shor,{\it The quantum channel capacity and coherent information}, lecture notes,
MSRI Workshop on Quantum Computation (2002); I. Devetak, IEEE
Trans. Inf. Th. {\bf 51}, 44 (2005); S. Lloyd, Phys. Rev. A {\bf
55}, 1613 (1997).
\bibitem{nonadditive} D.P. DiVincenzo, P.W. Shor, J.A. Smolin,
Phys. Rev. A {\bf 57}, 830 (1998).

\bibitem{DS} I. Devetak, P.W. Shor, quant-ph/0311131.
\bibitem{CG}  F. Caruso, V. Giovannetti, quant-ph/0603257.
\bibitem{WGPG} M.M. Wolf, D. Perez-Garcia, G. Giedke,
quant-ph/0606132.
\bibitem{GF} V. Giovannetti, R. Fazio, Phys. Rev. A {\bf 71}, 032314 (2005).
\bibitem{Jami} A. Jamiolkowski, Rep. Math. Phys., {\bf 3}, 275 (1972).
\bibitem{Ruskai} M. B. Ruskai, S. Szarek, E. Werner, Lin. Alg. Appl. {\bf 347}, 159
(2002).
\bibitem{cont} P. Horodecki, M.L. Nowakowski, quant-ph/0503070.
\bibitem{ED} D.P. DiVincenzo, P.W. Shor, J.A. Smolin, B.M. Terhal,
A.V. Thapliyal, Phys. Rev. A {\bf 61}, 062312 (2000); W. D\"ur,
J.I. Cirac, M. Lewenstein, D. Bruss, Phys. Rev. A {\bf 61}, 062313
(2000); P.W. Shor, J.A. Smolin, B.M. Terhal, Phys. Rev. Lett. {\bf
86}, 2681 (2001); T. Eggeling, K.G.H. Vollbrecht, R.F. Werner,
M.M. Wolf, Phys. Rev. Lett. {\bf 87}, 257902 (2001); K.G.H.
Vollbrecht, M.M. Wolf, Phys. Rev. Lett. {\bf 88}, 247901 (2002).
\bibitem{addQ} Let $m\geq m'$. By super-additivity of $Q$ together
with its assymptotic definition (i.e., $Q(T^{\otimes n})=n Q(T)$)
and the assumed additivity for $Q(T_1\otimes T_2)$ we have that
\begin{eqnarray*} Q\big(T_1^{\otimes m}\otimes T_2^{\otimes
m'}\big) &\leq& Q\big(T_1^{\otimes m}\otimes
T_2^{\otimes m}\big) - Q\big(T_2^{\otimes m-m'}\big)\\
&=& m Q\big(T_1\otimes T_2\big)-(m-m')Q(T_2)\\
&=& m Q(T_1) + m' Q(T_2)\;.
\end{eqnarray*} The converse inequality follows from
super-additivity.
\bibitem{SS} G. Smith, J.A. Smolin, quant-ph/0604107.
\bibitem{SSW} G. Smith, J.A. Smolin, A. Winter, quant-ph/0607039.

\end{thebibliography}
\end{document}